\documentclass[11pt,draftcls,onecolumn]{IEEEtran}

\makeatletter
\def\ps@headings{%
\def\@oddhead{\mbox{}\scriptsize\rightmark \hfil \thepage}%
\def\@evenhead{\scriptsize\thepage \hfil \leftmark\mbox{}}%
\def\@oddfoot{}%
\def\@evenfoot{}}
\makeatother

\usepackage{array, latexsym,times,amsfonts}
\usepackage{url}
\usepackage{algorithm,algorithmic}
\usepackage{epstopdf}

\usepackage{cite}
\usepackage{comment}

\begin{document}

\newif{\ifbibfile}
\bibfiletrue

\newcommand{\generation}{\mathcal{G}}
\newcommand{\n}{\mathcal{\rho}}
\newtheorem{theorem}{Theorem}[section]
\newtheorem{lemma}[theorem]{Lemma}
\newtheorem{proposition}[theorem]{Proposition}
\newtheorem{corolary}[theorem]{Corolary}
\newtheorem{ex}[theorem]{Example}
\newtheorem{definition}[theorem]{Definition}
\newtheorem{fact}[theorem]{Fact}
\newcommand{\auth}{\mbox{auth}}
\newcommand{\C}{\mathcal{C}}
\newcommand{\R}{\mathcal{R}}
\newcommand{\Se}{\mathcal{S}}
\newcommand{\group}{\mathcal{G}}
\newcommand{\field}{\mathcal{P}}
\newcommand{\pr}{{\rm Pr}}
\newcommand{\Simu}{\mathcal{S}}
\newcommand{\A}{\mathcal{A}}
\newcommand{\U}{\mathcal{U}}
\newcommand{\D}{\mathcal{D}}
\newcommand{\G}{\mathcal{G}}
\newcommand{\X}{\mathcal{X}}
\newcommand{\Y}{\mathcal{Y}}
\newcommand{\Z}{\mathbb{Z}}
\newcommand{\ID}{\mathrm{ID}}
\newcommand{\GA}{\mathcal{A}^G}
\newcommand{\hash}{\mathcal{H}}
\newcommand{\oracle}{\mathcal{O}}
\newcommand{\pairing}{\hat{e}}

\title{Password Protected Smart Card and Memory Stick Authentication 
Against Off-line Dictionary Attacks}

\author{Yongge Wang\\
UNC Charlotte, Charlotte, NC 28223, USA\\
\{yonwang\}@uncc.edu}

\maketitle

\begin{abstract}
In this paper, we study the security requirements for 
remote authentication with password protected smart card.
In recent years, several protocols for password-based authenticated key
exchange have been proposed. These protocols are used for the protection 
of password based authentication between a client and a remote server.
In this paper, we will focus on the password based authentication between
a smart card owner and smart card via an untrusted card reader. 
In a typical scenario, a smart card
owner inserts the smart card into an untrusted card reader and input
the password via the card reader in order for the smart card
to carry out the process of authentication with a remote server.
In this case, we want to guarantee that the card reader will not 
be able to impersonate the card owner in future without the smart card 
itself. Furthermore, the smart card could be stolen. If this happens,
we want the assurance that an adversary could not use the smart card
to impersonate the card owner even though
the sample space of passwords may be small enough to be enumerated by
an off-line adversary. At the end of this paper, we further extend
our results to credential storage on portable non-tamper resistant
storage devices such as USB memory sticks.
\end{abstract}

\noindent
{\bf ACM Computing Classification}: E.3 DATA ENCRYPTION;
E.4 CODING AND INFORMATION THEORY;
D.4.6 Security and Protection;
and K.6.5 Security and Protection.

\section{Introduction}
Numerous cryptographic protocols rely on passwords selected by users
(people) for strong authentication.  Since the users find it inconvenient
to remember long passwords, they typically select short easily-rememberable 
passwords. In these cases, the sample space of passwords may be small enough 
to be enumerated by an adversary thereby making the protocols vulnerable to
a \emph{dictionary} attack. It is desirable then to design password-based
protocols that resist off-line dictionary attacks (see, e.g., \cite{srp5}). 

The problem of password-based remote authentication protocols was first studied by 
Gong, Lomas, Needham, and Saltzer \cite{Gong93protectingpoorly} who used public-key
encryption to guard against off-line password-guessing attacks. 
In another very influential work \cite{Bellovin93augmentedencrypted}, Bellovin and Merritt 
introduced Encrypted Key Exchange (EKE), which became the basis
for many of the subsequent works in this area. These protocols
include SPEKE \cite{speke} and SRP \cite{srp3,srp5}. 
Other papers addressing the above protocol problem can be found in
\cite{Bellare00authenticatedkey,Black02cipherswith,Boyko00provablysecure,Bresson03securityproofs,Boyd01ellipticcurve}.
In models discussed in the above mentioned papers, we can assume that there is a trusted
client computer for the user to input her passwords. In a smart card based
authentication system, this assumption may no long be true. The smart card
reader could be malicious and may intercept the user input passwords. 
Furthermore, a smart card could be stolen and the adversary may launch
an off-line dictionary attack against the stolen smart card itself.
It is the goal for this paper to discuss the security models for
smart card based remote authentication and to design secure protocols within
these models.

In a practical deployment of smart card based authentication systems,
there may be other system requirements. For example, we may be 
required to use symmetric cipher based systems only or 
to use public key based systems.
Furthermore, the system may also require that the server 
store some validation data for each user or the server do 
not store any validation (this can be considered as identity based systems).
Furthermore, there may be other requirements such as user password
expiration and changes.

In the following, we use an example to show the challenges 
in the design of secure smart card based authentication protocols. 
A traditional way to store or transfer the secret key for each user is to
use a symmetric key cipher such as AES to encrypt user's long term secret key
with user's password and store the encrypted secret key on the user's 
smart card or USB memory card. This will not meet our security 
goals against off-line dictionary attacks. For example, in an RSA based 
public key cryptographic system, the public key is 
a pair of integers $(n,e)$ and the private key is an integer $d$.
With the above mentioned traditional approach, the smart card 
contains the value $AES_{\alpha}(d)$ in its tamper resistant 
memory space, where $\alpha$ is  the user's password. If such a card is stolen,
the adversary could feed a message (or challenge) $m$ to the smart card for 
a signature. The adversary needs to input a password in order for 
the smart card to generate a signature. The adversary will just pick one 
$\alpha'$ from her dictionary and ask the card to sign $m$. The card will 
return a signature $s'$ on $m$. If the guessed password $\alpha'$ is correct,
the adversary should be able to verify $s'$ on $m$. Otherwise, $s'$ could not 
be verified on $m$ with the public key $(n,e)$. Thus the attacker will remove
$\alpha'$ from the dictionary. Similar attacks work for 
Guillou-Quisquater (GQ), Fiat-Shamir, and Schnorr 
zero-knowledge identification schemes.

This example shows that the ``off-line'' dictionary attack 
in the stolen smart card environments is  different from the 
traditional client-server based off-line dictionary attacks. 
One potential approach to defeat this attack is to set a counter
in the smart card. That is, the smart card is allowed to sign a certain
number of messages, and then self-destroy it. However, the balance
between usability and security should be carefully considered here.
There have been quite a number of papers dealing with smart card 
based remote authentications 
(see, e.g., \cite{yalinchen,pass1,pass2,pass3,pass4,pass5,pass6}).  
However, most of them seem to be finding attacks on previous proposals 
and putting forward new ones without proper security justification
(or even a security model).

\section{Security models}
\label{security}

Halevi and Krawczyk \cite[Sections 2.2-2.3]{hk} introduced
a notion of security for remote authentication with memorable passwords. 
They provide a list of basic attacks that a password-based client-server 
protocol needs to guard against. Though these attacks are important
for password-based authentication, they are not sufficient for
password-protected smart card based remote authentication. 
In the following, we provide an extended list of attacks that a 
password-protected smart card based authentication protocol
needs to protect against. An ideal password-protected smart card 
protocol should be secure against these attacks and we will follow these
criteria when we discuss the security of password-protected smart card 
authentication protocols.
\begin{itemize}
\item {\bf Eavesdropping}. The attacker may observe the communications 
channel.
\item {\bf Replay}. The attacker records messages (either from the 
communication channels or from the card readers) she has observed
and re-sends them at a later time.
\item {\bf Man-in-the-middle}. The attacker intercepts the 
messages sent between the two parties (between users $\U$ and smart card $\C$
or between smart card $\C$ and servers $\Se$)
and replaces these with her own messages.  For example, if she 
sits between the user and the smart card, then
she could play the role of smart card in the messages
which it displays to the user on the card reader and at the same time 
plays the role of users to the smart card.
A special
man-in-the-middle attack is the \emph{small subgroup attack}. 
We illustrate this kind of attack by a small
example. Let $g$ be a generator of the group $\group$ of 
order $n=qt$ for some small $t>1$. In a standard
Diffie-Hellman key exchange protocol, the client $\C$ chooses a 
random $x$ and sends $g^x$ to the server 
$\Se$, then $\Se$ chooses a random $y$ and sends $g^y$ to $\C$.
The shared key between $\C$ and $\Se$ is $g^{xy}$.
Now assume that the attacker $\A$ intercepts $\C$'s message $g^x$,
replaces it with $g^{xq}$, and sends it to $\Se$. $\A$ also intercepts
$\Se$'s message $g^y$, replaces it with  $g^{yq}$, and sends it to $\C$.
In the end, both $\C$ and $\Se$ compute the shared key $g^{qxy}$.
Since $g^{qxy}$ lies in the subgroup of order $t$ of the group
generated by $g^q$, it takes on one of only $t$ possible values.
$\A$ can easily recover this  $g^{qxy}$ by an exhaustive search.
\item {\bf Impersonation}. The attacker impersonates the user 
(using a stolen smart card or a fake smart card) to an actual
card reader to authenticate to the remote server, 
impersonate a card reader to a user who inserts an authentic smart card, 
impersonate a card reader and a smart card (a stolen card or a fake card but 
without the actual user),
or impersonate the server to get some useful information.
\item {\bf Malicious card reader}. The attacker controls the card
reader and intercepts the smart card owner's input password.
Furthermore, the attacker controls all of the communications
between smart card and the card owner via the card reader,
and all of the communications between smart card and the remote server. 
For example, the attacker may launch a man in the middle attack 
between the smart card and smart card owner.
\item {\bf Stolen smart card}. The attacker steals the smart card
and impersonates the smart card owner to the remote server via a trusted 
or a malicious smart card reader. One exception
that we need to make in our security model is that we will not 
allow the attacker to control a malicious card reader to intercept the 
card owner's password and then to steal the smart card. There are three
kinds of attackers based on the stolen smart card scenario:
\begin{itemize}
\item {\em Smart card is tamper resistant with counter protection}. 
The attacker cannot read the sensitive information stored in the 
tamper resistant memory. Furthermore, the attacker may only 
issue a fixed amount of queries to the smart card to learn
useful information. The smart card will be self-destroyed if the
query number exceeds certain threshold (e.g., the GSM SIM card V2 or later
has this capability). 
\item {\em Smart card is tamper resistant without counter protection}. 
The attacker cannot read the sensitive information stored in the 
tamper resistant memory. However, the attacker may issue large amount of 
queries to the smart card to learn some useful information. For example, 
the attacker may set up a fake server and uses a malicious card reader 
to guess the potential password.
\item {\em Smart card is not tamper resistant}.
The attacker (with the card) may be able to break the tamper resistant 
protection of the smart card and read the sensitive information 
stored in the tamper resistant memory. In this case, the smart card will
look more like a USB memory stick that stores the user credential with
password protection. But still there is a difference here. In order 
for the user to use USB memory stick based credentials, the user needs
the access to a trusted computer to carry out the authentication. However,
one may assume that even if the smart card is not tamper resistant, it is 
not possible for a malicious card reader to read the sensitive information
on the card within a short time period (e.g., during the time that the 
card owner inserts the card into the card reader for an authentication). Thus 
no trusted card reader is required for this kind of smart card authentication.
\end{itemize}
\item {\bf Password-guessing}. The attacker is assumed to have 
access to a relatively small dictionary of words that likely includes the 
secret password $\alpha$. In an \emph{off-line attack}, the attacker 
records past communications and searches for a word in the dictionary that
is consistent with the recorded communications or carry out interaction with 
a stolen smart card without frequent server involvement (the attacker
may carry out one or two sessions with server involved and all other
activities without server involvement). In an \emph{on-line attack}, 
the attacker repeatedly picks a password from the dictionary and attempts 
to impersonate $\U$, $\C$, $\U$ and $\C$, or $\Se$. 
If the impersonation fails, the attacker
removes this password from the dictionary and tries again, using 
a different password. 
\item {\bf Partition attack}. The attacker records past communications,
then goes over the dictionary and deletes those words that are not  
consistent with the recorded communications from the dictionary. After 
several tries, the attacker's dictionary could become very small.
\end{itemize}
We now informally sketch the definition of security models.
We have three kinds of security models. 
\begin{enumerate}
\item {\bf Type I}. The attacker $\A$ is allowed to watch 
regular runs of the protocol between a smart card reader 
$\R$ (could be under the control of $\A$) and 
the server $\Se$, can actively communicate with $\R$ and $\Se$
in replay, impersonation, and man-in-the-middle attacks,
and can also actively control a smart card reader when the card  
owner inserts the smart card and inputs her password.
Furthermore, the attacker may steal the smart card from the user
(if this happens, we assume that the attacker has not observed the user
password from the previous runs of protocols)
and issue a large amount of queries to the smart card using 
a malicious card reader. However, we assume that the smart card is tamper
resistant and the attacker could not read the sensitive data from the 
smart card.
A protocol is said to be \emph{secure} in the presence of such an 
attacker if (i) whenever the server $\Se$ accepts an authentication
session with $\R$, it is the case that the actual user $\U$ 
did indeed insert her smart card into $\R$ and input the correct 
password in the authentication session; and 
(ii) whenever a smart card accepts an authentication
session with $\Se$, it is the case that $\Se$ did indeed participate in 
the authentication session and the user $\U$ did indeed input the correct
password.
\item {\bf Type II}. The capability of the attacker is the same as 
in the Type I model except that when the attacker steals the smart card,
it can only issue a fixed number of queries to the smart card using 
a malicious card reader. If the number of queries exceeds the threshold,
the smart card will be self-destroyed.
\item {\bf Type III}.  The capability of the attacker is the same as 
in the Type I model except that when the attacker steals the smart card,
it will be able to read all of the sensitive data out from the smart card.
But we will also assume that when a card owner inserts the card into
a malicious card reader for a session of authentication, the card reader
should not be able to read the information stored in the tamper resistant
section of the card. In another word, the smart card is not tamper resistant
only when the attacker can hold the card for a relatively long period by
herself. Another equivalent interpretation of this assumption is that
the attacker may not be able to intercept the password via the card reader
and read the information stored in the card at the same time.
\end{enumerate}

\section{Smart card based secure authentication and key agreement}
\subsection{Symmetric key based scheme: SSCA}
\label{symmetricID}
In this symmetric key based smart card authentication scheme SSCA, the server should choose  a master secret $\beta$ and protect it securely.
The Set up phase is as follows:
\begin{itemize}
\item For each user with identity $\C$ and password $\alpha$, the card maker (it knows the server's
master secret $\beta$) sets the card secret key as $K={\cal H}(\beta, \C)$ and 
 stores ${\cal K}={\cal E}_{\alpha}(K)$ in the tamper resistant
memory of the smart card, where ${\cal E}$ is a symmetric encryption algorithm such as AES and ${\cal H}$ is a hash algorithm such as SHA-2.
\end{itemize}

In the SSCA scheme, we assume that the smart card has the capability
to generate unpredictable random numbers. There are several ways for smart card to do so.
One of the typical approaches is to use hash algorithms and EPROM. 
In this approach, a random number is stored in the EPROM of the smart 
card when it is made. Each time, when a new random number is needed, 
the smart card reads the current random number in the EPROM 
and hash this random number with a secret key. 
Then it outputs this keyed hash output as the new random
number and replace the random number content in the EPROM with 
this new value. In order to keep protocol security, it is important 
for the smart card to erase all session information 
after each protocol run.  This will ensure that, in case the smart card 
is lost and the information within the tamper resistant memory is 
recovered by the attacker, the attacker should not able to recover 
any of the random numbers used in the previous runs of the protocols.
It should be noted that some smart card industry uses symmetric encryption
algorithms to generate random numbers. Due to the reversible operation
of symmetric ciphers, symmetric key based random number generation 
is not recommended for smart card implementation.

Each time when the user inserts her smart card into a card reader 
(which could be malicious), the card reader asks the user to 
input the password which will be forwarded the password 
to the smart card.

\begin{enumerate}
\item Using the provided password $\alpha$, the card decrypts 
$K={\cal D}_{\alpha}({\cal K})$. If the password is correct, 
the value should equal to ${\cal H}(\beta, \C)$. The card selects a random number $R_c$, computes 
$R_A={\cal E}_K(\C, R_c)$, and sends the pair
$(\C, R_A)$ to the card reader which will be forwarded to the server. 
\item The server recovers the value of $(\C, R_c)$ using the key $K={\cal H}(\beta, \C)$ 
and verifies that the identity $\C$ of the card  is correct. If the verification passes, 
the server selects a random number $R_s$, computes $R_B={\cal E}_K(\C, R_s)$, and sends 
$(\C, R_B, C_s)$ to the card reader which forwards it to the card. Here 
$C_s=\mbox{HMAC}_{sk}(\Se, \C, R_s, R_c)$ is the keyed message authentication tag on 
$(\Se, \C, R_s, R_c)$ under the key $sk={\cal H}(\C, \Se, R_c,R_s)$ and $\Se$ is the server 
identity string.
\item The card recovers the value of  $(\C, R_s)$ using the key $K={\cal H}(\beta, \C)$,
computes $sk={\cal H}(\C, \Se, R_c,R_s)$, and verifies the HMAC authentication tag $C_s$.
If the verification passes, it computes its own confirmation message as 
$C_c=\mbox{HMAC}_{sk}(\C, \Se, R_c, R_s)$ and sends $C_c$ to the server. 
The shared session key will be $sk$.
\item The server accepts the communication if the HMAC tag $C_c$ passes the verification.
\end{enumerate} 
The protocol SSCA message flows are shown in the Figure \ref{symprotocol}
\begin{figure}[htc]
\caption{Message flows in SSCA}
\label{symprotocol}
$$\begin{array}{ll}
\underline{\mbox{Card}}\longrightarrow  \underline{\mbox{Server}}: & 
    {\C, {\cal E}_K(\C, R_c)}\\
\underline{\mbox{Card}}\longleftarrow  \underline{\mbox{Server}}: & 
    {\cal E}_K(\C, R_s), C_s\\
\underline{\mbox{Card}}\longrightarrow  \underline{\mbox{Server}}: & 
    {C_c}
\end{array}$$
\end{figure}

In the following, we use heuristics to show that SSCA is secure in the Type I and 
Type II security models. If the underlying encryption scheme ${\cal E}$ and HMAC are secure,
then eavesdropping, replay, man-in-the-middle, impersonation,  
password-guessing, and partition attacks will learn nothing about the password since no
information of password is involved in these messages. Furthermore, a malicious 
card reader can intercept the password, but without the smart card itself, the attacker
will not be able to learn information about the secret key $K={\cal D}_{\alpha}({\cal K})$.
Thus the attacker will not be able to impersonate the server or the card owner.
When the attacker steals the smart card (but she has not controlled a cart reader to
intercept the card owner password in the past), she may be able to insert the card
into a malicious card reader and let the card to run the protocols with a fake server polynomial
many times. In these protocol runs, the attacker could input guessed password $\alpha'$.
The smart card will  output $(\C, {\cal E}_{K'}(\C, R_c))$ where $K'={\cal D}_{\alpha'}({\cal K})$.
Since the attacker has no access to the actual server (this is an off-line attack), the attacker
can not verify whether the output $({\C, {\cal E}_{K'}(\C, R_c)})$ is in correct format. Thus the attacker has
no way to verify whether the guessed password $\alpha'$ is correct. In a summary, the
protocol is secure in the Type I and Type II security models.

The protocol SSCA is not secure in the Type III security model. Assume
that the attacker has observed a previous valid run 
of the protocol (but did not see the password) before steal the smart card. 
For each guessed password $\alpha'$, the attacker computes a potential 
key $K'=\D_{\alpha'}({\cal K})$. If this key $K'$ is not consistent
with the observed confirmation messages in the previous run of the protocol, 
the attacker could remove $\alpha'$ from the password list. Otherwise,
it guessed the correct password. 

If we revise the attacker's capability in Type III model by restricting
the attacker from observing any valid runs of the protocol
before she steals the smart card, we get a new security model which we will call Type III$'$ model.
We can show that the protocol SSCA is secure in the Type  III$'$ model. The heuristics is that 
for an attacker with access to the value ${\cal K}={\cal E}_{\alpha}(K)$, he will not 
be able to verify whether a guessed password is valid off-line. For example, for each guessed
password $\alpha'$, she can compute  $K'={\cal D}_{\alpha'}({\cal K})$. But she has no idea
whether $K'$ is the valid secret key without on-line interaction with the server. Thus the protocol
is secure in the Type  III$'$ security model.

{\bf Remarks}: Modification of the protocol may be necessary for certain
applications. For example, if the card identification string $\C$ 
itself needs to be protected (e.g., it is the credit card number), 
then one certainly does not want to transfer the identification string
$\C$ along with the message in a clear channel.

\subsection{Public key based scheme: PSCAb}
\label{wangkesection}

In this section, we introduce a public key based smart card authentication 
scheme with bilinear groups: PSCAb, it is based on the identity based key agreement protocol 
from IEEE 1363.3 \cite{ieee1363,wangke}.

In the following, we first briefly describe the bilinear maps
and bilinear map groups. 
\begin{enumerate}
\item $G$ and $G_1$ are two (multiplicative) cyclic groups of
prime order $q$.
\item $g$ is a generator of $G$.
\item $\pairing:G\times G\rightarrow G_1$ is a bilinear map.
\end{enumerate}
A bilinear map is a map $\pairing:G\times G\rightarrow G_1$ 
with the following properties:
\begin{enumerate}
\item bilinear: for all $g_1,g_2\in G$, and $x,y\in Z$, we have 
$\pairing(g_1^x,g_2^y)=\pairing(g_1,g_2)^{xy}$.
\item non-degenerate: $\pairing(g,g)\not=1$.
\end{enumerate}
We say that $G$ is a bilinear group if the group action in $G$ can be 
computed efficiently and there exists a group $G_1$ and an efficiently
computable bilinear map $\pairing:G\times G\rightarrow G_1$ as above.
For convenience, throughout the paper, we view both $G$ and $G_1$ 
as multiplicative groups though the concrete implementation of $G$ 
could be additive elliptic curve groups.

Let $k$ be the security parameter given to the setup algorithm 
and $\mathcal{IG}$ be a bilinear group parameter generator.
We present the scheme by describing the
three algorithms: {\bf Setup}, {\bf Extract}, and {\bf Exchange}. 

\vskip 3pt
\noindent
{\bf Setup}: For the input $k\in Z^{+}$, 
the algorithm proceeds as follows:
\begin{enumerate}
\item Run $\mathcal{IG}$ on $k$ to generate a bilinear group
$G_\n=\{G,G_1,\pairing\}$ and the prime order $q$ of the two groups 
$G$ and $G_1$. 
Choose a random generator $g\in G$.
\item Pick a random master secret $\beta\in  Z^*_q$.
\item Choose cryptographic hash functions $\hash:\{0,1\}^*\rightarrow G$
and $\pi:G\times G\rightarrow Z_q^*$. In the security analysis, 
we view $\hash$ and $\pi$ as random oracles. 
\end{enumerate}
The system parameter is $\langle q, g, G, G_1, \pairing, 
\hash,\pi\rangle$ and the master secret key is $\beta$.

\vskip 3pt
\noindent
{\bf Extract}: For a given identification 
string $\C\in \{0,1\}^*$, the algorithm 
computes a generator $g_{\C}=\hash(\C)\in G$, and sets the private key 
$d_{\C}=g_{\C}^{\beta}$ where $\beta$ is the master secret key. The algorithm will
further compute $g_{\Se}=\hash(\Se)\in G$ where $\Se$ is the server 
identity string, 
and store the value $\left(\C, g_{\Se}, d_{\C}^{\hash(\alpha)}\right)$
in the tamper resistant smart card.

\vskip 3pt
\noindent
{\bf Exchange}: The algorithm proceeds as follows.
\begin{enumerate}
\item The card selects $x\in_R Z_q^*$, 
computes $R_A=g_{\C}^{x}$, and sends it to the Server via the card reader.
\item  The Server selects $y\in_R Z_q^*$, 
computes $R_B=g_{\Se}^{y}$, and sends it to the card.
\item The card computes
$s_A=\pi(R_A,R_B)$, $s_B=\pi(R_B,R_A)$, 
${\cal H}(\alpha)^{-1} \mbox{ mod } q$, and 
the shared session key $sk$ as 
$$\pairing(g_{\C}, g_{\Se})^{(x+s_A)(y+s_B)\beta}=\pairing\left(d_{\C}^{(x+s_A){\cal H}(\alpha){\cal H}(\alpha)^{-1}}, g_{\Se}^{s_B}\cdot R_B\right).$$
\item The card computes $K_1=\hash(sk,1)$ and sends 
$C_{\C}=\mbox{HMAC}_{K_1}(\C,\Se,R_A,R_B)$ to the server.
\item The server computes $s_A=\pi(R_A,R_B)$, $s_B=\pi(R_B,R_A)$ and  
the shared session key $sk$ as 
$$\pairing(g_{\C}, g_{\Se})^{(x+s_A)(y+s_B)\beta}=
\pairing\left(g_{\C}^{s_A}\cdot 
R_A, g_{\Se}^{(y+s_B)\beta}\right).$$
\item The server verifies whether $C_{\C}$ is correct. If the verification passes. The server
computes $K_1=\hash(sk,1)$ and sends 
$C_{\Se}=\mbox{HMAC}_{K_1}(\Se,\C,R_B,R_A)$ to the card.
\item The card verifies the value of $\C_{\Se}$.
\end{enumerate}

The protocol PSCAb message flows are shown in the Figure \ref{symprotocolpscab}
\begin{figure}[htc]
\caption{Message flows in PSCAb}
\label{symprotocolpscab}
$$\begin{array}{ll}
\underline{\mbox{Card}}\longrightarrow  \underline{\mbox{Server}}: &  g_{\C}^x\\
\underline{\mbox{Card}}\longleftarrow  \underline{\mbox{Server}}: &  g_{\Se}^y\\
\underline{\mbox{Card}}\longrightarrow  \underline{\mbox{Server}}: &  C_{\C}\\
\underline{\mbox{Card}}\longleftarrow  \underline{\mbox{Server}}: &  C_{\Se}\\
\end{array}$$
\end{figure}

In the following, we use heuristics to show that PSCAb is secure in the Type I, 
Type II, and Type III security models. The security of the underlying identity based key agreement
protocol IDAK \cite{wangke} (it is called Wang Key Agreement protocol in \cite{ieee1363})
is proved in \cite{wangke}. Furthermore, the eavesdropping, replay, man-in-the-middle, impersonation,  
password-guessing, and partition attacks will learn nothing about the password since no
information of password is involved in these messages. Furthermore, these attackers will 
learn nothing about the private keys $d_{\C}$ and $\beta$ based on the proofs in \cite{wangke}. 
For an attacker with access to the information $d_{\C}^{\hash(\alpha)}$ (the attacker may read
this information from the stolen smart card), she may impersonate the card owner to 
interact with the server. Since the attacker could only compute the value $sk^{\hash(\alpha)}$,
it will not be able to generate the confirmation message $C_\C$. Thus the server will not 
send the server confirmation message back to the attacker. In another word, the attacker will 
get no useful information for an off-line password guessing attack.
Furthermore, even if the attacker has observed previous valid protocol runs, it will not help
the attacker due to the fact that the smart card does not contain any information 
of the session values $x$ of the previous protocols runs.

{\bf Remarks}:
In the protocol PSCAb, it is important to have the card to send the confirmation message to 
the server first. Otherwise, the protocol will not be secure in the Type III security model. 
Now we assume that the server
sends the first confirmation message and 
we present an attack in the following with this protocol variant. After the attacker obtains
the value $d_{\C}^{{\cal H}(\alpha)}$ from the smart card, she 
could impersonate the user by sending the vale $R_A$ to the server.
Though the attacker may not be able to compute the shared secret $sk$ with 
the server, it will be able to compute 
$$sk^{{\cal H}(\alpha)}=
\pairing\left(d_{\C}^{(x+s_A){\cal H}(\alpha)}, 
g_{\Se}^{s_B}\cdot R_B\right).$$
For each guessed password $\alpha'$, the attacker computes a potential 
secret $sk'=sk^{{\cal H}(\alpha) {\cal H}(\alpha')^{-1}}$. 
If $sk'$ is not consistent with the confirmation message $C_{\Se}$ from the 
server, the attacker could remove $\alpha'$ from the password list. Otherwise,
it guessed the correct password.

\subsection{Public key based scheme with password validation data at server: PSCAV}
\label{pscavsection}

In previous sections, we discussed two protocols SSCA and PSCAb that the server does not 
store any password validation data. In this section, we discuss some protocols that server
needs to store password validation data for each card. One of the 
disadvantages of this kind of protocols is that if the card owner wants
to change her password, the server has to be involved.

It should be noted that the password based remote authentication protocols
that have been specified in the IEEE 1363.2 \cite{ieee1363} are not 
secure in our models. The major reason is that the only secure 
credential that a client owns is the password. If the smart card 
owner inputs her password on an untrusted card reader, the card 
reader could just record the password and impersonates the client
to the server without the smart card in future.

Before we present our scheme PSCAV, we briefly note that the protocol PSCAb 
in Section \ref{wangkesection} can be easily modified to be a password protected
smart card authentication scheme that the server stores user password validation data.
In Section \ref{wangkesection}, the identity string for each user is computed as 
$g_{\C}=\hash(\C)\in G$. For protocols with password validation data, we can 
use a different way to compute the identity strings. In particular, assume that the user 
$\U$ has a password $\alpha$, then the identity string for the user will be computed 
as $g_{\C}=\hash(\C, \alpha)\in G$
and the private key for the user will be $d_{\C}=g_{\C}^{\beta}$ where 
$\beta$ is the master secret key. The value 
$\left(\C, g_{\Se}, d_{\C}^{{\cal H}(\alpha)}\right)$ will be stored 
in the tamper resistant smart card, and the value  $g_{\C}$ will be stored in the 
server database for this user. The remaining of protocol runs the same
as in  Section \ref{wangkesection}. We can call the above mentioned protocol as PSCAbV

Now we begin to describe our main protocol PSCAV for this section. 
Assume that the server has a master secret $\beta$.
For each user with password $\alpha$, let the user specific generator be
$g_\C={\cal H}_1(\C,\alpha, \beta)$, the value $g_\C^{{\cal H}_2(\alpha)}$ is stored on 
the smart card, where $\hash_2$ is another independent hash function. 
The value $g_\C={\cal H}_1(\C,\alpha, \beta)$ will be stored in the server database for this user. 
The remaining of protocol runs as follows:
\begin{enumerate}
\item The card selects random $x$ and sends $R_A=g_\C^x$ to the server.
\item Server selects random $y$ and sends $R_B=g_\C^y$ to the card.
\item The card computes $u=\hash(\C,\Se,R_A,R_B)$ where 
$\Se$ is the server identity string, $sk=g_\C^{y(x+u\alpha)}$,
and sends $C_c=\hash(sk,\C,\Se,R_A,R_B)$ to the server
\item The server verifies that the $C_c$ is correct. If the verification passes, server 
computes $u=\hash(\C,\Se,R_A,R_B)$, $sk=g_\C^{y(x+u\alpha)}$,
and sends $C_s=\hash(sk,\Se,\C,R_B,R_A)$ to the card.
\end{enumerate}

The protocol PSCAV message flows are the same as for the PSCAb protocol 
message in the Figure  \ref{symprotocolpscab} (but with different 
interpretation for the variables in the figure).

In the following, we use heuristics to show that PSCAV is secure in the Type I, 
Type II, and Type III security models. For the PSCAV protocol,
the eavesdropping, replay, man-in-the-middle, card (client) impersonation,  
password-guessing, and partition attacks will learn nothing about the password 
due to the hardness of the Diffie-Hellman problem. For the attacker that carries out a server
impersonation attack, it will receive the value $R_A$, and send a random $R_B$ to the card.
The attacker will then receive the card confirmation message $C_\C$. The attacker may not 
launch an off-line dictionary attack on these information since for each guessed password
$\alpha'$, it has no way to generate a session key $sk'$ due to the hardness of the Diffie-Hellman problem. 
For an attacker with access to the information $\hash_1(\C,\alpha,\beta)^{\hash_2(\alpha)}$ 
(the attacker may read this information from the stolen smart card), she may impersonate 
the card owner to interact with the server. The attacker may send a random $R_A$ to the server
which could be based on $\hash_1(\C,\alpha,\beta)^{\hash_2(\alpha)}$, and receives 
a value $R_B$ from the server. But it cannot compute the session key $sk$ based
on these information. Thus it could not send the confirmation message $C_\C$ to the 
server. Thus the server will not  send the server confirmation message back to the attacker. 
In another word, the attacker will get no useful information for an off-line password guessing attack.
Furthermore, even if the attacker has observed previous valid protocol runs, it will not help
the attacker due to the fact that the smart card does not contain any information 
of the session values $x$ of the previous protocols runs.

{\bf Remarks}: The attack described at the end of Section
\ref{wangkesection} could be used to show that,
in the protocol PSCAV, it is important to have the card to send
the confirmation message to the server first. Otherwise, the protocol 
will not be secure in the Type III security model.

\section{Remote Authentication with password protected portable memory sticks}
In this section, we investigate the scenario that the user stores her 
private key on a USB memory stick. Our goal is that if the memory stick is 
lost, then the adversary will not be able to mount an off-line dictionary 
attack to impersonate the legitimate user. Since a memory stick will not have its own
CPU, the owner has to insert the memory stick into a trusted computer (otherwise,
the malicious computer could just intercept the password and copy the content
on the memory sticks and impersonates the owner in future).
Thus the security model for this kind of protocols are different 
from the Type I, II, III models that we have discussed in Section \ref{security}, but are closely
related to the Type III model. 
Specifically, we will have the following Type IV model for portable memory sticks.
\begin{itemize}
\item {\bf Type IV}.  The attacker $\A$ is allowed to watch 
regular runs of the protocol between a client
$\C$ (together with the memory stick and a trusted computer) and 
the server $\Se$, can actively communicate with $\C$ and $\Se$
in replay, impersonation, and man-in-the-middle attacks,
and can also steal the memory stick from the user and read the content in the
memory stick. A protocol is said to be \emph{secure} in the presence of such an 
attacker if (i) whenever the server $\Se$ accepts an authentication
session with $\C$, it is the case that the actual user $\U$ 
did indeed insert her memory stick into a computer $\C$ and input the correct 
password in the authentication session; and 
(ii) whenever a client $\C$  accepts an authentication
session with $\Se$, it is the case that $\Se$ did indeed participate in 
the authentication session and the user $\U$ did indeed input the correct
password.
\end{itemize}

Heuristics could be used to show that the protocols PSCAb from section
\ref{wangkesection} and PSCAV from section \ref{pscavsection} are secure
in the Type IV model. Based on similar heuristics, it can also be shown
that the protocol SSCA from section \ref{symmetricID} will be 
secure in modified Type IV model in the same way as the modified 
security model Type III discussed in  section \ref{symmetricID}.
In another word, if we revise the attacker's capability in Type IV 
model by restricting the attacker from observing any valid runs of 
the protocol before she steals the memory stick, 
we get a new security model which we will call Type iV$'$ model.
It can be shown that the protocol SSCA is secure in the Type  IV$'$ model. 

Though we have showed that all of our three protocols PSCAb, PSCAV,
and SSCA have similar characteristics in both the security 
models Type III and Type IV, it is an open question whether 
the security models Type III and Type IV are equivalent. In another 
word, we do not know whether there are protocols that are secure 
in Type III model but not secure in Type IV model (or vice versa).

\bibliographystyle{plain}
\ifbibfile
\bibliography{yonggewang}
\fi

\end{document}

\end{document}